\documentclass[apjl]{emulateapj-rtx4}
\usepackage{natbib}
\usepackage{amsmath}
\usepackage{amssymb}
\usepackage{amsthm}
\usepackage{bm}
\usepackage{dcolumn}
\usepackage{graphicx}
\usepackage{subfigure}
\usepackage{color}

\begin{document}

\title{Progenitor models of the electromagnetic transient associated with the short GRB 130603B}
\author{Kenta Hotokezaka$^1$}
\author{Koutarou Kyutoku$^2$}
\author{Masaomi Tanaka$^3$}
\author{Kenta Kiuchi$^4$}
\author{Yuichiro Sekiguchi$^4$}
\author{Masaru Shibata$^4$}
\author{Shinya Wanajo$^3$}

\affiliation{
$^1$Department of Physics,~Kyoto~University,~Kyoto~606-8502,~Japan\\
$^2$Department of Physics, University of Wisconsin-Milwaukee, P.O. Box 413,
Milwaukee, Wisconsin 53201, USA\\
$^3$National Astronomical Observatory of Japan, Mitaka, Tokyo, Japan\\
$^4$Yukawa Institute for Theoretical Physics, Kyoto University,
Kyoto 606-8502, Japan \\
}

\begin{abstract}
An electromagnetic transient powered by the radioactive decay of {\it r}-process elements, a so-called
kilonova/macronova, is one of the possible observable consequences of compact
binary mergers including at least one neutron star. Recent observations strongly
suggest the first discovery of the electromagnetic transient, which is associated with the short
GRB 130603B. We explore a possible progenitor of this event combining the
numerical-relativity simulations and radiative transfer simulations of
the dynamical ejecta of binary neutron star and black hole - neutron star
mergers. We show that the ejecta models within a realistic parameter range
consistently reproduce the observed near-infrared excess. We also show
that the soft equation of state models for binary neutron star mergers and the stiff
equation of state models for
black hole - neutron star mergers are favored to reproduce the observed luminosity.

\end{abstract}

\maketitle

\section{Introduction}
Mergers of compact binaries such as binary neutron stars~(NS-NSs) and black
hole-neutron star binaries~(BH-NSs)\footnote{A compact binary implies an NS-NS
or BH-NS binary throughout this Letter.} are the candidates for progenitors of
short-hard gamma-ray bursts~(GRBs)~\citep{paczynski1986ApJL,goodman1986ApJL,eichler1989Nature}.
As an observable consequence of compact
binary mergers, an electromagnetic transient powered by radioactive decay of the heavy elements
produced in their ejecta (so-called kilonova/macronova)
was proposed by \citet{li1998ApJ}. Recently, a strong evidence of the electromagnetic transient
associated with the short GRB 130603B was observed by {\it Hubble Space Telescope} 
\citep{tanvir2013Nature,berger2013ApJ}.
The observed near-infrared excess is largely consistent with the radioactively powered emission models of NS-NS
merger ejecta~\citep{kasen2013ApJ,barnes2013ApJ,tanaka2013ApJ,grossman2013}
\footnote{See \citet{jin2013ApJ} for an alternative interpretation, but the non-detection of
late-time radio emission of GRB 130603B provides a strong constraint on it~\citep{fong2013}.}. This could be
a direct evidence for the compact binary merger scenario of short GRBs.

Characteristics of the radioactively powered transient are determined primarily by {\it r}-process elements
which are expected to be produced in the ejecta of compact binary mergers because of their neutron-richness
\citep{lattimer1974ApJ,symbalisty1982ApL,meyer1989ApJ,freiburghaus1999ApJ,goriely2011ApJ,korobkin2012MNRAS,
rosswog2013MNRAS,bauswein2013ApJa}. 
The decay of the {\it r}-process elements heats the ejecta and gives rise to an electromagnetic transient
emission~\citep{metzger2010MNRAS,roberts2011ApJ,goriely2011ApJ,grossman2013}.
The {\it r}-process elements play also an important role as the dominant opacity source of the
ejecta. 
The bound-bound transitions of partially ionized {\it r}-process elements
significantly enhance the opacity of the ejecta in ultraviolet to near-infrared wavelengths~\citep{kasen2013ApJ,tanaka2013ApJ}.

The brightness and time scale of an electromagnetic transient depend also on the
global properties of the ejecta such as the mass,
expansion velocity, and morphology. Using the observed data of the near-infrared excess, 
\cite{berger2013ApJ} estimated the ejecta mass and expansion velocity under
the assumption of spherically expanding ejecta.
These properties of the ejecta depend on the type of the progenitor~(NS-NS or BH-NS),
on the parameters of the progenitor models such as mass and spin of the two objects, and on the equation of state (EOS)
of neutron-star matter. Recent numerical simulations of compact binary
mergers have explored the ejecta properties~\citep{rosswog1999A&A,oechslin2007A&A,hotokezaka2013PRDa,
bauswein2013ApJa,rosswog2013RSPTA,foucart2013PRD,lovelace2013CQG,kyutoku2013PRD,deaton2013}.  
By these numerical studies, we can estimate the dependences of the ejecta
properties on the parameters of the progenitor models and on EOSs.
Therefore, combining the observed data of the near-infrared excess and the results of numerical simulations,
it is possible to constrain the progenitor models of
the short GRB 130603B.   

\begin{figure*}[t]
\begin{tabular}{l l}
\rotatebox{0}{\includegraphics[width=85mm,clip]{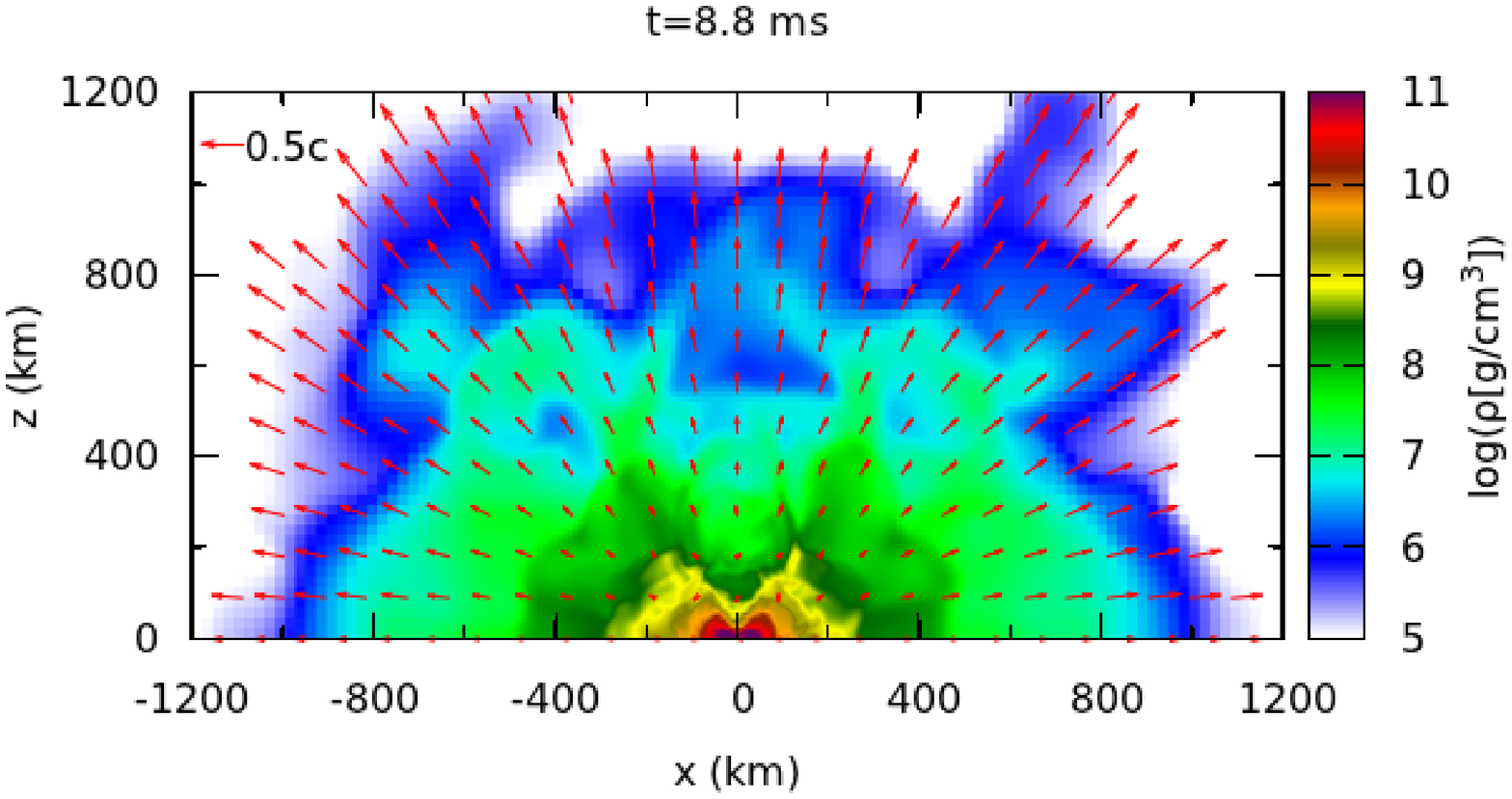}}
\rotatebox{0}{\includegraphics[width=85mm,clip]{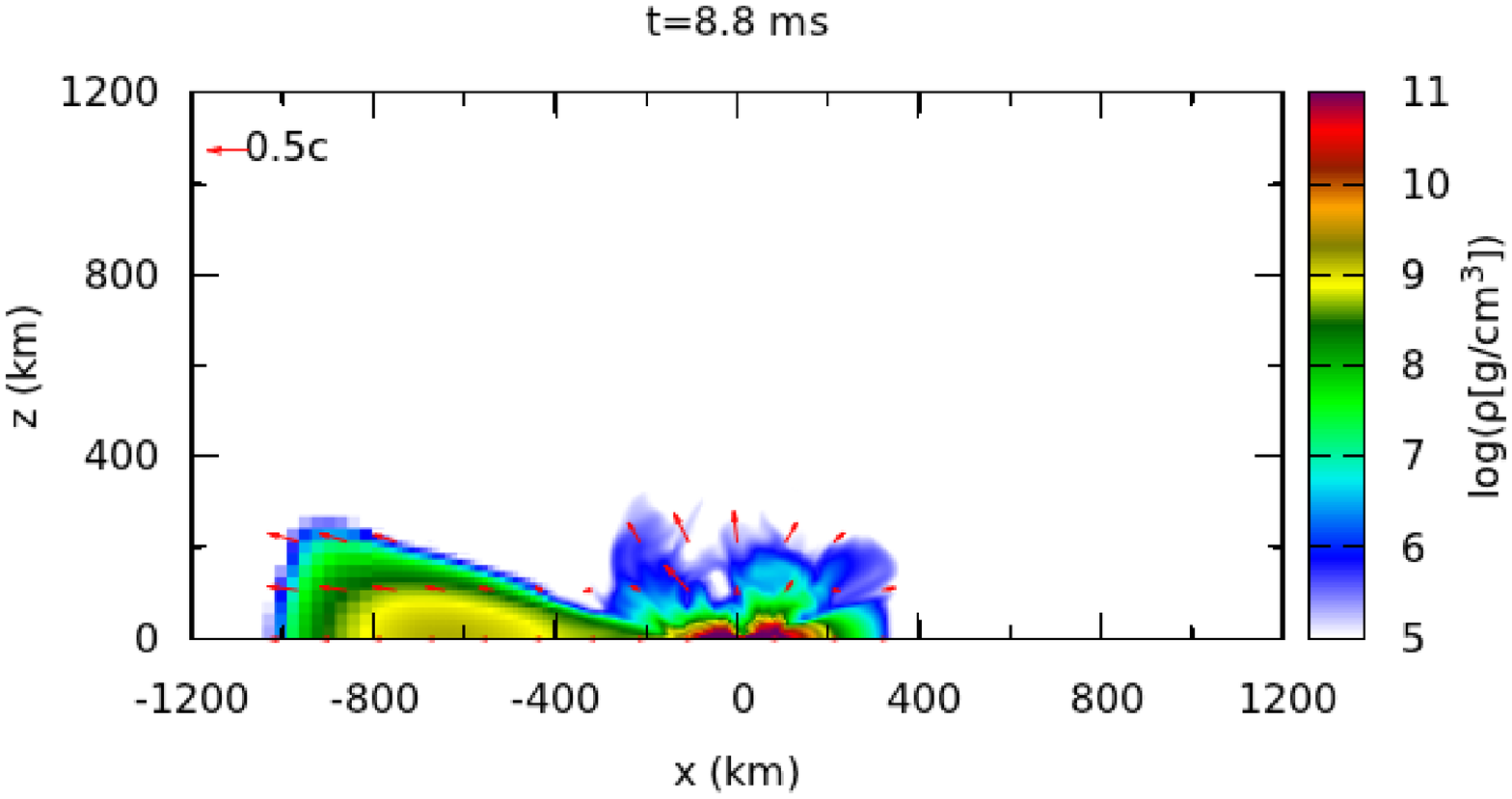}}
\end{tabular}
\caption{The rest-mass density profiles on the meridional plane for the NS-NS (SLy,~$M_{\rm{tot}}=2.7M_{\odot},~Q=1.0$)
and BH-NS (H4,~$Q=3,~\chi=0.75$) models at $8.8$~ms after the onset of the merger. }
\label{fig1}
\end{figure*}

In this Letter, we explore possible progenitor models of the electromagnetic transient
associated with the short GRB 130603B using our numerical-relativity simulations and radiative
transfer simulations of the dynamical ejecta of compact binary mergers.
We focus in particular on its dependence on the EOS of neutron-star matter.

\section{Mass ejection of compact binary mergers}

The density and velocity structures of the ejecta of compact
binary mergers are determined by the numerical-relativity simulation
using SACRA code~\citep{yamamoto2008PRD}. 
We follow the dynamical
ejecta with the numerical-relativity simulation until the head of
the ejecta reaches $\simeq 1000$~km~(see \citealt{hotokezaka2013PRDa} and
\citealt{kyutoku2013PRD} for details). After that, the density and velocity 
structures of the ejecta are modeled assuming homologous expansion~\citep{rosswog2013}. 
For the simulations, we employ
a piecewise polytropic EOS with which the cold EOSs of neutron-star
matter are well fitted~\citep{read2009PRD}. For systematic studies
of the dependence of mass ejection on the cold EOSs of neutron-star matter,
we consider five cold EOSs: APR4~\citep{akmal1998PRC} and SLy~\citep{douchin2001A&A}
as soft EOSs, ALF2~\citep{alford2005ApJ} as a moderate EOS, and
H4~\citep{glendenning1991PRL,lackey2006PRD} and MS1~\citep{muller1996NuPhA}
as stiff EOSs\footnote{In this Letter, `soft' and `stiff' EOSs mean those which
reproduce the radii $R_{1.35}\leq 12$~km and $R_{1.35} \geq 13.5$~km, respectively.
Here $R_{1.35}$ is the radius of a cold, spherical neutron star with the gravitational mass $1.35M_{\odot}$.
For all the EOSs, the maximum masses of spherical neutron stars are larger than $\simeq 2M_{\odot}$.}. 

For NS-NS mergers, we choose the total gravitational mass of the binary $M_{\rm{tot}}=2.6M_{\odot}
- 2.8M_{\odot}$ and the mass ratio\footnote{The mass ratio is defined
by $Q=m_{1}/m_{2}$ with $m_{1} \geq m_{2}$, where $m_{1}$ and $m_{2}$ are the component masses
of a binary.} $Q=1.0-1.25$.
For BH-NS mergers, the gravitational mass
of the neutron star $M_{\rm{NS}}$ is fixed to be $1.35M_{\odot}$ and the mass ratio is chosen
to be $Q=3-7$. The nondimensional spin parameter of the black hole $\chi$ is
chosen as $\chi = 0.75$. We also perform the simulations for $Q=7$ and $\chi=0.5$.  
These parameters, ejecta masses $M_{\rm{ej}}$, and averaged ejecta velocities $\langle v_{\rm{ej}}\rangle /c$ of the progenitor
models are summarized in Table~\ref{tab1}.

The morphologies of the ejecta for NS-NS and BH-NS mergers are compared
in Fig.~\ref{fig1}. This figure plots the profiles of the density and velocity
fields at $8.8$~ms after the onset of the merger. 
Note that the ejecta velocities are in the small range between $\sim 0.1c$ and $\sim 0.3c$, 
because those are roughly equal to the escape velocities
from the neutron stars, which do not depend sensitively on the progenitor models.
However, the total mass and morphology of the ejecta depend sensitively on the progenitor models. 
In the following, we briefly summarize these properties of the NS-NS and BH-NS ejecta. 

\begin{table*}[ht]
\caption[]{
  Parameters of the progenitor models and their ejecta properties.}
\begin{center}
 \begin{tabular}{lccccccc} \hline \hline
 \textrm{EOS} & type & $R_{1.35}$&$M_{\rm{tot}}/M_{\odot}$ &$Q$& $\chi$ &$M_{\rm{ej}}/10^{-2}M_{\odot}$ & $\langle v_{\rm{ej}}\rangle /c$  \\ \hline
APR4 & NS-NS & 11.1 &2.6--2.9 & 1.0--1.25 & -- & 0.01--1.4 & 0.22--0.27\\
SLy & NS-NS & 11.4 & 2.6--2.8 & 1.0--1.25 & --  & 0.8--2   & 0.2--0.26\\ 
ALF2 & NS-NS & 12.4 &2.6--2.8 & 1.0--1.25 & --  & 0.15--0.55 & 0.22--0.24\\
H4 & NS-NS& 13.6 &2.6--2.8 & 1.0--1.25 & -- & 0.03--0.4  & 0.18--0.26\\
MS1 & NS-NS& 14.4 &2.6--2.8 & 1.0--1.25 & --  & 0.06--0.35 & 0.18--0.2 \\ \hline 
APR4 & BH-NS & 11.1 &5.4--10.8 & 3.0--7.0  & 0.75 & 0.05--1 & 0.23--0.27\\
ALF2 & BH-NS & 12.4 &5.4--10.8 & 3.0--7.0 & 0.75  & 2.0--4.0 & 0.25--0.29\\
H4 & BH-NS & 13.6 &5.4--10.8 & 3.0--7.0 & 0.75  & 4.0--5.0  & 0.24--0.29\\
MS1 & BH-NS & 14.4 &5.4--10.8 & 3.0--7.0 & 0.75  & 6.5--8.0 & 0.25--0.3\\ 
\hline
APR4 & BH-NS & 11.1 &5.4--10.8 & 7.0  & 0.5 & $\lesssim 10^{-4}$ & --\\
ALF2 & BH-NS & 12.4 &5.4--10.8 & 7.0 & 0.5  & 0.02  & 0.27\\
H4 & BH-NS & 13.6 &5.4--10.8 & 7.0 & 0.5  & 0.3 & 0.29\\
MS1 & BH-NS & 14.4 &5.4--10.8 & 7.0 & 0.5  & 1.7 & 0.3\\ 
\hline \hline
\label{tab1}
\end{tabular}
\end{center}
\end{table*}

{\it NS-NS ejecta.}
As shown in Fig.~\ref{fig1}, the NS-NS ejecta have  spheroidal shape rather than a
torus or a disk irrespective of $Q$ and EOS as long as a hypermassive neutron star
is formed after the merger. The reason is as follows. The origin of the ejecta for
NS-NS mergers can be divided into two parts: the contact interface of two neutron
stars at the collision and the tidal tails formed during an early
stage of the merger. At the contact interface, the
kinetic energy of the approaching velocities of the two stars is converted into
thermal energy through shock heating. The heated matter at the contact
interface expands into low-density region. As a result, the shocked matter
can escape even toward the rotational axis and the ejecta shape becomes spheroidal. 
By contrast, the tidal tail component is asymmetric and the ejecta is distributed
near the equatorial plane.

\begin{figure*}[t]
\begin{tabular}{l l}
\hspace{-3mm}
\rotatebox{0}{\includegraphics[width=90mm,clip]{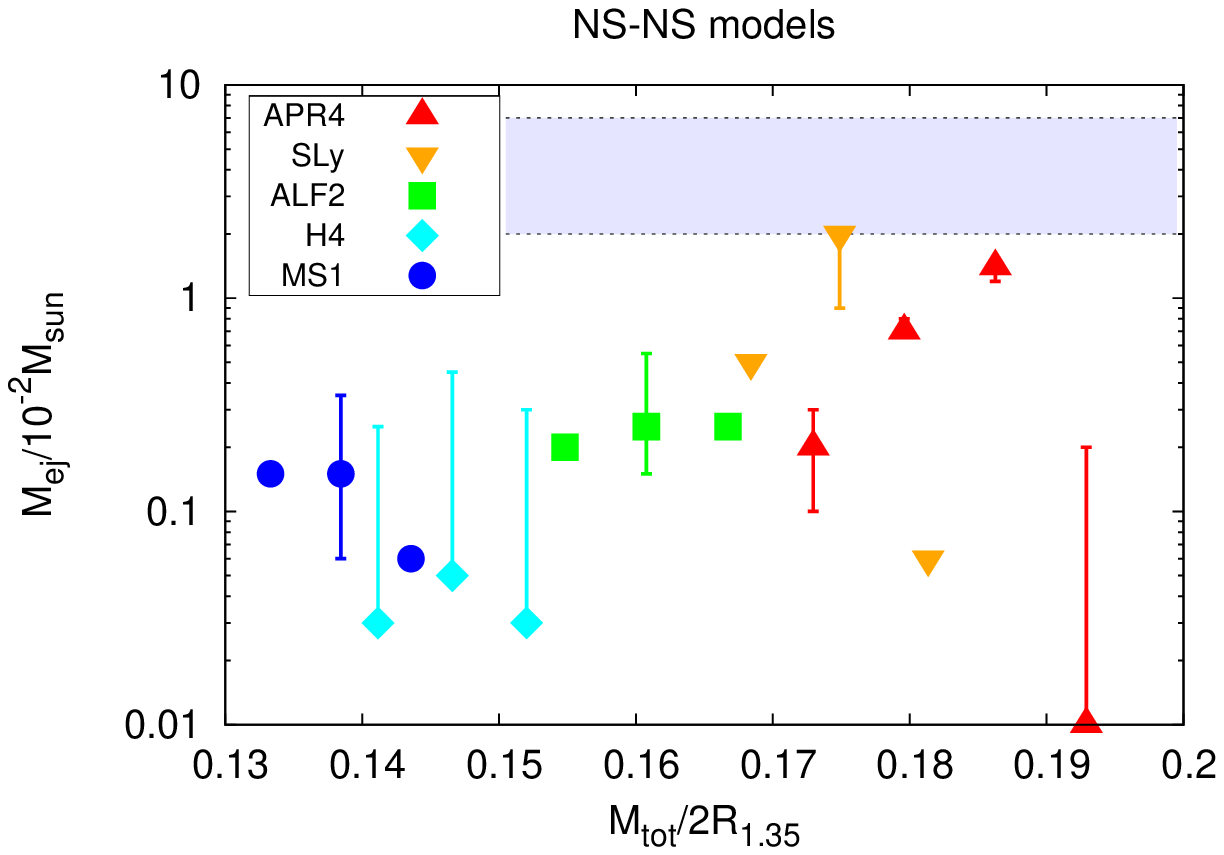}}
\rotatebox{0}{\includegraphics[width=90mm,clip]{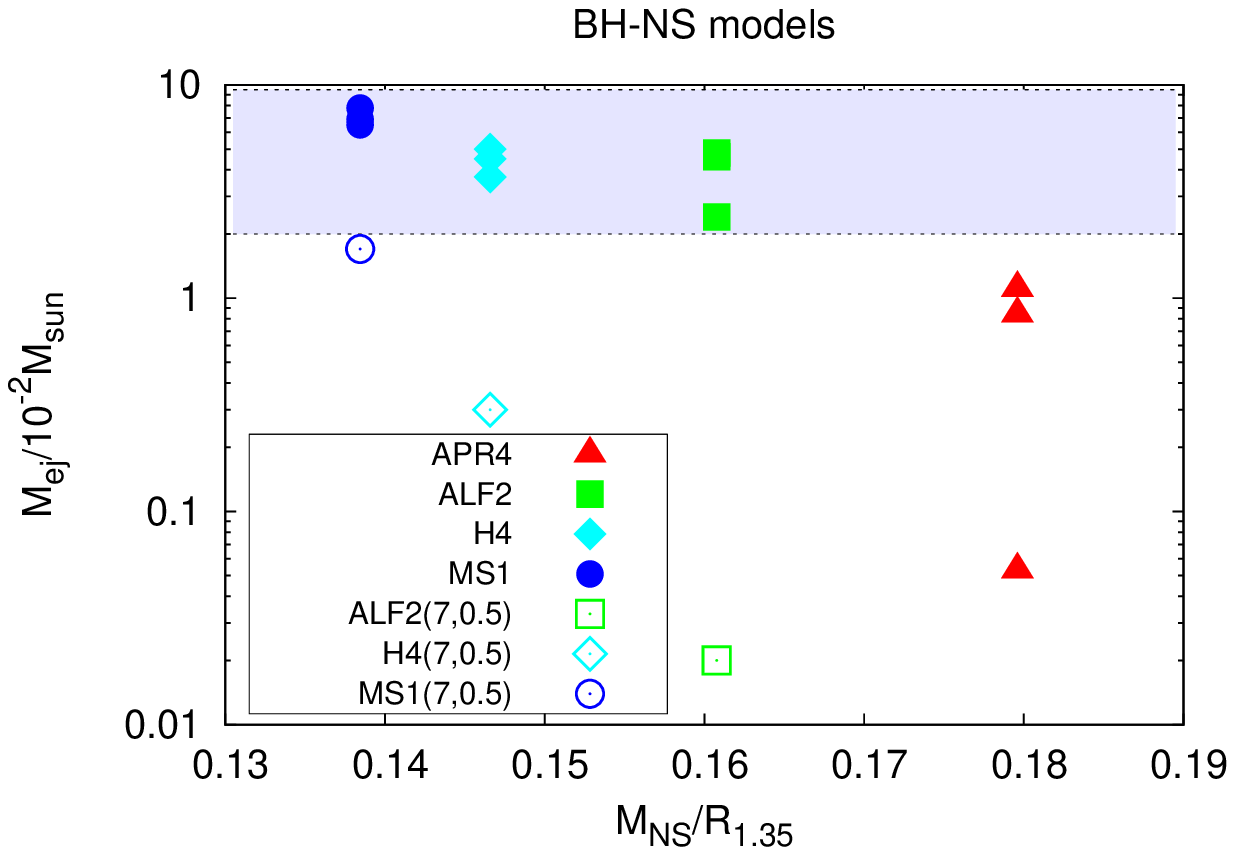}}
\end{tabular}
\caption{ Ejecta masses as a function of the compactness of the neutron star, which is defined by $GM_{\rm{tot}}/2R_{1.35}c^{2}$
and $GM_{\rm{NS}}/R_{1.35}c^{2}$ for NS-NS and BH-NS models, respectively.
Left panel: NS-NS models. Each point shows the ejecta mass for the equal mass cases.
Error bars denote the dispersion of the ejecta masses due to the various mass ratios.
Right panel: BH-NS models. The filled and open symbols correspond to the models with $(Q,\chi)=(3$--$7,0.75)$ 
and $(7,0.5)$, respectively. The blue shaded region in each panel shows the allowed
ejecta masses to reproduce the observed near-infrared excess of GRB 130603B,
$0.02\lesssim M_{\rm{ej}}/M_{\odot} \lesssim 0.07$ and $0.02 \lesssim M_{\rm{ej}}/M_{\odot} \lesssim 0.1$ for
NS-NS and BH-NS models, respectively. The lower and upper bounds are imposed by the hypothetical high- and
low-heating models, respevtively.}
\label{fig2}
\end{figure*}

Numerical simulations of NS-NS mergers show that the total amount of ejecta
is in a range $10^{-4}$--$10^{-2}M_{\odot}$ depending on $M_{\rm{tot}}$, $Q$,
and EOSs~(see the left panel of Fig.~\ref{fig2}).
The more compact neutron star models with soft EOSs produce the larger amounts of ejecta,
because the impact velocities and subsequent shock heating effects at merger
are larger.
More specifically, the amounts of the ejecta are
\begin{eqnarray}
10^{-4} \lesssim M_{\rm{ej}}/M_{\odot} \lesssim 2\times 10^{-2} ~~~(\rm{soft~EOSs}),\nonumber \\ 
10^{-4} \lesssim M_{\rm{ej}}/M_{\odot} \lesssim 5\times 10^{-3} ~~~(\rm{stiff~EOSs}).
\end{eqnarray}

\cite{bauswein2013ApJa} also show a similar dependence of the ejecta masses
on the EOSs and $M_{\rm{ej}}\lesssim 0.01M_{\odot}$ for stiff EOS models.
According to these results, it is worthy to note that
the ejecta masses of the stiff EOS models are likely to be at most $0.01M_{\odot}$.
 
The dependence of the ejecta mass on the total mass of the binary is rather complicated
as shown in the left panel of Fig.~\ref{fig2}. 
The ejecta mass increases basically with increasing $M_{\rm{tot}}$ as long as
a hypermassive neutron star with lifetime $\gtrsim 10$~ms is formed after the merger.
More massive NS-NS mergers result in hypermassive neutron stars with lifetime $\lesssim 10$~ms
or in black holes. For such a case, the ejecta mass decreases with increasing $M_{\rm{tot}}$
because of the shorter duration for the mass ejection.

{\it BH-NS ejecta.}
Tidal disruption of a neutron star results in the anisotropic mass
ejection for a BH-NS merger~\citep{kyutoku2013PRD}. As a result, the 
ejecta concentrate near the binary orbital plane as shown in the
right panel of Fig.~\ref{fig1}, and its shape is like a disk or crescent.

The amounts of the ejecta for BH-NS models are smaller for
the more compact neutron star models with the fixed values of $\chi$
and $Q$ as shown in the right panel of Fig.~\ref{fig2}. This is because
the tidal disruption occurs in a less significant manner for the
more compact neutron star models.
This dependence of the BH-NS ejecta on the
compactness of neutron stars is opposite to the NS-NS ejecta.

More specifically, the amounts of the ejecta are
\begin{eqnarray}
5\times10^{-4} \lesssim M_{\rm{ej}}/M_{\odot} \lesssim 10^{-2} ~~~(\rm{soft~EOSs}), \nonumber \\ 
4\times10^{-2} \lesssim M_{\rm{ej}}/M_{\odot} \lesssim 7\times10^{-2} ~~~(\rm{stiff~EOSs}),
\end{eqnarray}
for $\chi=0.75$ and $3\leq Q\leq 7$. 
For $\chi=0.5$, the ejecta mass is smaller than that for $\chi=0.75$.
Only the stiff EOS models can produce large amounts of ejecta more than $0.01M_{\odot}$
for $\chi=0.5$ and $Q=7$.

Both for NS-NS and BH-NS merger models, winds driven by neutrino/viscous/nuclear-recombination
heating or magnetic field from the central object might
provide ejecta in addition to the dynamical ejecta~\citep{dessart2009ApJ,wanajo2012ApJ,kiuchi2012PRD,fernandez2013MNRAS}.
However, it is not easy to estimate the amount of the wind ejecta,
because it depends strongly on the condition of the remnant formed 
after the merger. In this Letter, we focus only on the dynamical ejecta. 

\section{Radiative transfer simulations for the ejecta}

For the NS-NS and BH-NS merger models described in Section 2,
we perform radiative transfer simulations
to obtain the light curves of the radioactively powered emission from the ejecta
using the three-dimensional, time-dependent, multi-frequency Monte Carlo
radiative transfer code~\citep{tanaka2013ApJ}.
For a given density structure of the ejecta and
elemental abundances,
this code computes the emission in the ultraviolet, optical,
and near-infrared wavelength ranges by taking into account
the detailed opacities of $r$-process elements.
In this Letter, we include $r$-process elements with $Z \geq 40$
assuming the solar abundance ratios 
by~\citet{simmerer2004ApJ}.
More details of the radiative transfer simulations
are described in \citet{tanaka2013ApJ,tanaka2013b}.

The heating rate from the radioactive decay of $r$-process elements is one of
the important ingredients of the radiative transfer simulations. 
As
a fiducial-heating model, we employ the heating rate computed with the
abundance distribution that reproduces the solar $r$-process pattern~(see
Tanaka et al. 2013 for more detail). Heating is due to $\beta$-decays
only, which increase the atomic numbers from the neutron-rich region
toward the $\beta$-stability line without changing the mass number $A$. This heating
rate is in reasonable agreement with those from previous nucleosynthesis
calculations~\citep{metzger2010MNRAS, goriely2011ApJ, grossman2013}
except for the first several seconds.

We note that there could exist quantitative uncertainties
in the heating rate as well as in the opacities. 
As an exapmle, the heating rate would be about a factor 2 higher if the
$r$-process abundances of $A\sim 130$ (or those produced with the electron fraction
of $Y_{\rm{e}}\sim 0.2$) were dominant in the ejecta~\citep{metzger2010MNRAS,grossman2013}.
To take into account such uncertainties, we also consider the cases in which the light curves of mergers
are twice and half as 
luminous (high- and low-heating models; only explicitly shown for the NS-NS models in Fig.~3)
as those computed with the fiducial-heating model.

\section{Light curves and possible progenitor models}

\begin{figure*}[t]
\begin{tabular}{l l}
\rotatebox{0}{\includegraphics[width=90mm,clip]{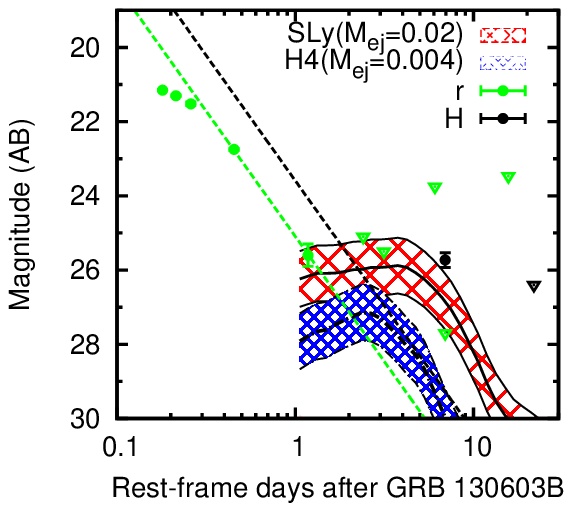}}
\rotatebox{0}{\includegraphics[width=90mm,clip]{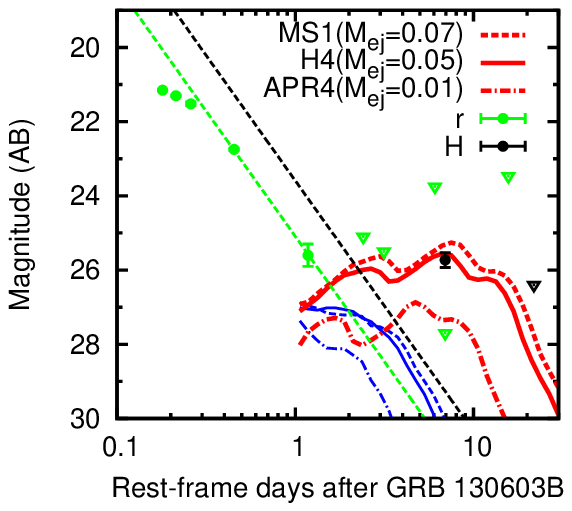}}
\end{tabular}
\caption{Predicted light curves for NS-NS and BH-NS models.
Left panel: NS-NS models. The dashed, solid, and dot-dashed curves show the
{\it H}-band light curves for the models: SLy~($Q=1.0,~M_{\rm{ej}}=0.02M_{\odot}$),
H4~($Q=1.25,~M_{\rm{ej}}=4\times 10^{-3}M_{\odot}$),
respectively. The total mass of the progenitor is fixed to be $2.7M_{\odot}$.
The upper, middle, and lower curves for each model correspond to the high-,
fiducial- and low-heating models. Right panel: BH-NS models. The dashed, solid, and
dot-dashed curves show the models MS1~($M_{\rm{ej}}=0.07M_{\odot}$), H4~($M_{\rm{ej}}=0.05M_{\odot}$),
and APR4~($M_{\rm{ej}}=0.01M_{\odot}$), respectively. Here only the fiducial-heating models are shown.
The thin and thick lines denote
the {\it r} and {\it H}-band light curves. Here we set $(Q,~\chi)=(3,~0.75)$.
The observed data and the light curves of the afterglow model
of GRB 130603B in {\it r} and {\it H}-band are plotted~\citep{tanvir2013Nature}.
The observed point in {\it r}-band at 1~days after the GRB is consistent with
the afterglow model. The key observations for an electromagnetic transient
are the observed {\it H}-band data at 7~days after the GRB,
which exceed the {\it H}-band light curve of the afterglow model,
and the upper limit in {\it H}-band at 22~days after the GRB.
These data suggest the existence of an electromagnetic transient associated with GRB 130603B. 
}
\label{fig3}
\end{figure*}

The computed light curves and observed data in {\it r} and {\it H}-band are compared
in Fig.~\ref{fig3}.  The left panel of Fig.~\ref{fig3} shows the light curves of
the NS-NS merger models SLy~($Q=1.0$,~$M_{\rm{ej}}=0.02M_{\odot}$)
and H4~($Q=1.25$,~$M_{\rm{ej}}=4\times 10^{-3} M_{\odot}$)
for reference. Here the total mass of the binary is chosen to be $M_{\rm{tot}}=2.7M_{\odot}$.
We plot three light curves derived with the fiducial-~(the middle curves),
high-~(the upper curves), and low-heating models~(the lower curves).
We expect that the realistic light curves may lie within the shaded regions.
For the NS-NS models, the computed {\it r}-band light curves are fainter than 30~mag.
The right panel of Fig.~\ref{fig3} shows the light curves of the BH-NS merger models,
MS1~($M_{\rm{ej}}=0.07M_{\odot}$), H4~($M_{\rm{ej}}=0.05M_{\odot}$), and APR4~($M_{\rm{ej}}=0.01M_{\odot}$)
with $(Q,~\chi)=(3,~0.75)$. For these cases, we employ the fiducial-heating model.
Note that the {\it r}-band light curves of the BH-NS models reach $\sim 27$~mag,
which implies that the light curves of the BH-NS models are bluer than those of the NS-NS models.
This is because the energy from the radioactive decay is deposited to a small volume for
the BH-NS models~(see \citet{tanaka2013b} for details).

We now translate these results into the progenitor models,
such as mass ratio, black hole spin, and EOS. 

{\it NS-NS models.}
The NS-NS models for GRB 130603B should have ejecta of mass
$\gtrsim 0.02M_{\odot}$. This is consistent with that derived by~\citet{berger2013ApJ}.
This value strongly constrains the NS-NS models because the amount of
the ejecta is at most $\sim 0.02M_{\odot}$ for an NS-NS merger within
the plausible mass range of the observed NS-NS systems~\citep{ozel2012ApJ}.
Specifically, as shown in the left panel of Fig.~\ref{fig2}, 
such a large amount of ejecta can be obtained only for the
soft EOS models in which a hypermassive
neutron star with lifetime $\gtrsim 10$~ms is formed after the merger. For the stiff EOS models,
the amount of the ejecta is at most $4\times 10^{-3}M_{\odot}$.
Thus we conclude that the ejecta of the  NS-NS models with soft EOSs
($R_{1.35}\lesssim 12$~km) are favored as the progenitor of GRB 130603B.

{\it BH-NS models.}
The observed data in the {\it H}-band is consistent with the BH-NS models which
produce the ejecta of $\sim 0.05M_{\odot}$ in our fiducial-heating model.
Such a large amount of ejecta can be obtained with only the
stiff EOSs~($R_{1.35}\gtrsim 13.5$~km) for the case of $\chi=0.75$ and 
$3\leq Q\leq 7$ as shown in the right panel of Fig.~\ref{fig2}. For the soft EOS models, the total amount of ejecta
reaches only $0.01M_{\odot}$ as long as $\chi\leq 0.75$, which hardly reproduces the observed near-infrared excess.
Thus the models with stiff EOSs are favored for the BH-NS merger models as long as with
$0.5 \leq \chi \leq 0.75$ and $3 \leq Q \leq 7$ as the progenitor
model of GRB 130603B. It is worthy to note that any BH-NS models with $\chi\leq 0.5$ and $Q \geq 7$
are unlikely to reproduce the observed near-infrared excess.

\section{Conclusion and Discussion}

We explored possible progenitor models of the electromagnetic transient associated with the short GRB 130603B.
This electromagnetic transient may have been powered by the radioactive decay of
{\it r}-process elements, so called kilonova/macronova.
We analyzed the dynamical ejecta of NS-NS and BH-NS mergers
for the progenitor models of this event. For computing the expected light curves,
we carried out the radiative transfer simulations using the density and velocity
structures obtained from the numerical-relativity simulations with several total masses,
mass ratios, and EOSs. Depending on these quantities, the total amount of ejecta mass
varies by orders of magnitude $10^{-4}M_{\odot}$ to $10^{-2}M_{\odot}$ for the NS-NS models
and $10^{-5}M_{\odot}$ to $10^{-1}M_{\odot}$ for the BH-NS models.
The expected light curves for the BH-NS models are bluer than those for the NS-NS models
due to the morphology effects.

For both NS-NS and BH-NS models, we found that there are progenitor models that can reproduce
the observed near-infrared excess within the realistic parameter ranges.
Specifically, the observed data suggest that the required ejecta mass is at least $\sim 0.02M_{\odot}$
for NS-NS mergers. For BH-NS mergers, the required ejecta mass would be $\sim 0.02$--$0.1M_{\odot}$
taking into account the uncertainty in the heating rate and opacities.
These values are consistent with the results of
a spherically expanding ejecta model~\citep{berger2013ApJ}. Such a large amount of material is ejected
when a hypermassive neutron star with its lifetime $\gtrsim 10$~ms is formed after the merger for
the NS-NS models and when the neutron star is tidally disrupted for the BH-NS models. 
For these cases, the merger results in a spinning black hole surrounded by a massive torus $\sim 0.1M_{\odot}$.
Such a remnant could have been the central engine of GRB 130603B.  
 
We constrained the progenitor models of GRB 130603B, which should produce the required amount of ejecta.  
We found that the soft EOS models are favored for NS-NS models.
For BH-NS models with the mass ratio $3 \leq Q \leq 7$ and the nondimensional
spin parameter of the black hole $0.5 \leq \chi \leq 0.75$, the stiff EOS models
are favored. For $\chi \leq 0.5$, any BH-NS models with $Q \geq 7$ are unlikely to produce
the required amount of ejecta.
In future, the observations of gravitational waves from compact binary mergers within $\sim 200$~Mpc
will provide the masses of the binaries and their types. Combining the observations of the gravitational-wave
and electromagnetic signals, it will be possible to constrain more stringently the progenitor models,
in particular EOSs, of such events.

\begin{acknowledgments}

We thank T. Nakamura for fruitful discussions.  This work was supported
by Grant-in-Aid for Scientific Research (21340051, 23224004, 23750160, 24244028, 
24740117, 24740163, 25103510, 25105508, 25103512)
by Grant-in-Aid for Scientific Research on Innovative Area
(20105004, 25103515), and HPCI Strategic Program of Japanese
MEXT. This work was partly supported by ``Joint Usage/Research Center
for Interdisciplinary Large-scale Information Infrastructures'' in
Japan. The work of Kyutoku is supported by JSPS Postdoctoral Fellowship for
Research Abroad. The work of Hotokezaka was supported by the Grant-in-Aid for
JSPS fellow Grant Number 24-1772.
This work was in part done in the 
longterm workshop on {\em Gravitational Waves and Numerical Relativity} held at the Yukawa
Institute for Theoretical Physics, Kyoto University in May and June 2013.   

\end{acknowledgments}

\end{document}